\documentclass[amssymb,twocolumn,10pt,prl]{revtex4}
\usepackage{dcolumn}
\usepackage{bm}

\begin{document} 
\noindent
{\large \bf Comment on ``Bell's Theorem without Inequalities and without 
Probabilities for Two Observers"} \\

In this Comment we show that Cabello's argument~$[1]$ which proves
 the nonlocal feature of any classical model of quantum mechanics based on 
 Einstein-Podolsky-Rosen (EPR) criterion of elements of reality, must involve 
 at least four distant observers rather than the two employed by 
 the author.
Therefore we claim that Cabello's proof is not only more complicated than 
 Hardy's argument~$[2]$, but it is also less cheap (in terms of physical 
 resources exploited) than the Greenberger-Horne-Zeilinger (GHZ)
 argument~$[3]$, which needs 
 only three particles and three observers to reach the same conclusions.  

First of all, let us briefly review Cabello's argument.
Starting from a four-particle quantum state $\vert \psi \rangle_{1234}$, 
 tensor product of two singlet states, he infers the existence of elements 
 of reality, as defined in~$[4]$, for a certain
 number of spin-observables of all the particles involved.
The values of these elements of reality, denoted with
 $v(A_{i}),v(a_{i}),v(B_{j})$ and $v(b_{j})$, are binded to satisfy certain
 algebraic constraints which he proves cannot be all fulfilled simultaneously.
The conclusion of~$[1]$ is that the hypothesis of locality, 
 i.e. the fact that elements of reality objectively possessed by a physical
 system cannot be altered instantaneously at a distance, must no longer hold 
 true in any deterministic completion of quantum mechanics.
It is worth noticing that Cabello is interested only in analyzing the particular
 case of possible deterministic completions of quantum mechanics, in which the 
 values objectively possessed by the observables are deduced 
 via an EPR criterion of reality and by counterfactual reasonings, rather than 
 postulated from the very beginning.

Let us now show that its argument, as it stands, is not correct since 
 two other observers, besides Alice and Bob, located in two spacelike
 separated spatial locations, are needed.
Suppose that Bob has, already and in a separate way, ascertained 
 the existence of the elements of reality $v(B_{2})$ and $v(B_{4})$
 and suppose that Alice, given $\vert \psi \rangle_{1234}$,
 measures the observable $A_{1}A_{3}$ finding the result $+1$. 
The state after the measurement process collapses to $\vert \tilde{\psi}
 \rangle_{1234}= 1/\sqrt{2} \,[\,\vert 0101 \rangle + 
 \vert 1010\rangle\,]_{1234}$.
Now, even if elements of reality for observables $B_{2}$ and $B_{4}$ 
 have already been deduced to exist (but are still unknown), there is no 
 way to infer the validity of the relation $v(B_{2})\!=\!v(B_{4})$ without
 invoking hidden variables but to resort to an EPR reasoning involving a 
 spacelike separation between the particles $2$ and $4$.  
In fact, given the state $\vert \tilde{\psi}\rangle$, we cannot reject the 
 possibility that,
 for example, $v(B_{2})\!\!=+1$ and $v(B_{4})\!\!=-1$ only on the ground that
 measurements of $B_{2}$ and $B_{4}$ always give equal outcomes for two 
 reasons: i) state $\vert \tilde{\psi}\rangle$ does not contain the maximal 
 specification of the properties of the system and ii) if the 
 particles are not spacelike separated we cannot dismiss the possibility
 that those measurement outcomes are equal due to a spooky causal influence 
 caused by the measurement processes performed on each particle. 
Being constrained to use EPR criterion of reality only, particles
 $2$ and $4$ must be in distant locations in order to deduce, in the usual
 way of reasoning, the fact that $v(B_{2})\!\!=\!\!v(B_{4})$.
A similar remark can be developed for the couple of observables $A_{1}$ and
 $a_{3}$, involving particles $1$ and $3$, once the  outcome $B_{2}b_{4}=+1$ 
 has been already found.
 
Therefore Cabello's proof needs a total number of four distant 
 observers (one for each particle) in order to be definitely correct. 
A spacelike separation between all of them is required in order to prevent that 
 instantaneous causal influences can be responsible for the validity of 
 relations like $v(B_{2})=v(B_{4})$ or $v(A_{1})=v(a_{3})$.
 
So, Cabello's proof of non-locality cannot be considered 
 {\em cheaper} (in terms of the number of observers and particles involved) 
 than GHZ argument: three spin one-half particles and three distant
 observers are (up to now) the minimum number of resources necessary to 
 exhibit the ``non-locality without inequalities" proof of any
 deterministic completion of quantum mechanics, working in $100\%$ of the
 runs.
 
Finally, we raise one last important remark on the necessity 
 of performing a real experiment confirming Cabello's argument.
Validity of equations from (3) to (11) of the paper~$[1]$ cannot be obviously 
 simultaneously verified for i) they involve non-compatible measurements and 
 ii) the mere act of ascertaining the joint outcomes of equation~(11) of~$[1]$
 invalidate the predictions of the other equations as a consequence
 of the wave function collapse. 
Thus, (3) to (11) are counterfactual properties that cannot be tested in a 
 single experiment but only separately, as clearly stated in the paper~$[1]$. 
But these experiments are in principle quite superfluous since they merely 
 mean to test one more time the validity of the predictions of quantum 
 mechanics, which have been already confirmed beyond every reasonable doubt. \\
 The author thanks D.Mauro for valuable comments.\\
  
 \noindent Luca Marinatto (marinatto@ts.infn.it) \\
 International Centre for Theoretical Physics\\ 
 ``Abdus Salam", Trieste, Italy \\

\noindent $[1]$ A.Cabello, {\it Phys.Rev.Lett.}, {\bf 86}, 1911 (2001).\\
$[2]$ L.Hardy, {\it Phys.Rev.Lett.}, {\bf 71}, 1665 (1993). \\
$[3]$ D.M.Greenberger, M.A.Horne, and A.Zeilinger, in {\it Bell's Theorem,
 Quantum Theory and Conceptions of the Universe}, ed. M.Kafatos
 (Kluwer, Dordrecht, 1989). \\
$[4]$ A.Einstein, B.Podolsky, and N.Rosen, {\it Phys.Rev.}, {\bf 47}, 777 
(1935).

\end{document}